\documentstyle[prl,aps,epsf,multicol]{revtex}

\begin{document}

\title{Relaxation of non-order parameter field in directed Ising systems}

\author{Heung Sik Park and Hyunggyu Park }

\address{Department of Physics, Inha University, Inchon 402-751, Korea }
\date{\today}
\maketitle
\begin{abstract}
We investigate the effect of initial conditions on the
dynamic exponents of the interacting monomer-monomer model with
infinitely many absorbing states in one dimension.
This model exhibits a directed Ising (DI) type transition from an active
phase into an absorbing phase. In case of the directed percolation
universality class, it has been reported that the non-order parameter
as well as the order parameter exhibits critical fluctuations,
relaxing algebraically to its natural value with the same scaling
exponents. We numerically confirm that this is also valid for the
DI universality class. We also observe continuously varying dynamic
exponents with a linear dependence on the non-order parameter
initial density.
\end{abstract}

\pacs{PACS Numbers: 64.60.-i, 05.40.+j, 82.20.Mj, 05.70.Ln}

\begin{multicols}{2}

Various kinds of nonequilibrium lattice models
exhibiting absorbing phase transitions have been studied
extensively during last few decades\cite{Dickman}.
Two distinct types of absorbing phase
transitions have been identified in one dimension: the directed
percolation (DP) and directed Ising (DI) universality class. Most
models have been found to belong to the DP class, which involves
typically a single absorbing state or multiple absorbing states
without any symmetry. DI-type critical behavior appears in
models with two equivalent absorbing states or two equivalent
group of absorbing states \cite{Park,Hwang}.

Recently, the question has been addressed whether one can construct
initial states that affect the {\em entire} temporal evolution of
nonequilibrium systems including asymptotic dynamics.
Typical examples are the systems that
display an absorbing phase transition from an active phase into
an absorbing phase with infinitely many absorbing (IMA) states.
As in the ordinary order-disorder phase transition models,
the order parameter takes a nonzero value in the active phase
and vanishes in the absorbing phase. The IMA states can be
characterized by a {\em non-order} parameter.
In the steady state, the non-order parameter approaches algebraically
to a {\em natural} value at criticality.

Nonuniversal dynamic properties
have been reported in various kinds of DP-type nonequilibrium lattice
models with IMA states \cite{Mendes,Marques,Jensen}.
The order-parameter dynamic exponents vary continuously with the initial
conditions characterized by the non-order parameter density.
Moreover, at criticality, the non-order parameter density also exhibits critical
fluctuations, relaxing algebraically to its natural value with the same
dynamic exponents \cite{Odor}. The critical
relaxation of the non-order parameter field in
DP-type systems with IMA states was
recently examined from a field theoretical (Langevin equation)
approach\cite{Grass1,Munoz}. The evolution equation for the order parameter
is non-Markovian and includes a temporal memory term due to
the non-order parameter field. This non-Markovian term
is known to be responsible for the nonuniversal dynamic
exponents.

In this paper, we introduce an interacting monomer-monomer model with IMA states
(IMA-IMA model) that belongs to the DI universality class and
investigate the effects of the non-order parameter field in one dimension.
Using numerical simulations, we show that the non-order parameter
plays the same role in DI systems as well as DP systems.

The IMA-IMM model is an interacting monomer-monomer model with two different
species of monomers, namely $A$ and $B$. Monomers $A$ and $B$ are
selected with probability $p$ and $1-p$ respectively. Monomers
can adsorb at a randomly selected vacant site with unit probability for $A$
and $1-r_B$ for $B$ respectively.  An adsorption attempt of a monomer is rejected when
both sites adjacent to a selected vacant site are occupied or
at least one adjacent site is occupied with a monomer of the same
species. A nearest neighbor $AB$ pair reacts and desorbs immediately from the lattice.
\begin{eqnarray}
A + V & \stackrel{p}{\longrightarrow}& A_s  \nonumber, \\
B + V &\stackrel{(1-p)(1-r_b)}{\longrightarrow}& B_s,  \\
A_s + B_s &\longrightarrow& \emptyset. \nonumber 
\end{eqnarray}
Here the subscript $s$ denotes adsorbed particles, $V$ denotes a
vacant site where the adsorption attempt is allowed.

This model has infinitely many absorbing states. Any
configuration without a nearest neighbor pair of vacant sites is
one of the absorbing states. The number of absorbing states
diverges exponentially with system size $L$ as $2^{L/2}$. The
absorbing states can be divided into two equivalent groups;
odd-site occupied and even-site occupied groups. Clearly, these
two groups of absorbing states have one-to-one correspondence and
the Ising ($Z_2$) symmetry in between. So we expect that this
model exhibits the DI-type absorbing phase transition.

We locate the critical line in the $r_b-p$ phase diagram by dynamic Monte Carlo
simulations \cite{PP}. For small values of $r_b$, the system is always absorbing.
As $r_b$ increases, a window of the active phase appears and divides the absorbing phase by
$A$-dominated and $B$-dominated absorbing phases. For example,
as $p$ increases along the $r_b=0.9$ line, the system undergoes two continuous phase transitions
from the $B$-dominated absorbing phase into the active phase at $p_{c1}=0.100(3)$
and finally into the $A$-dominated absorbing phase at $p_{c2}=0.5058(7)$.
We numerically confirm that both transitions are of the DI type
as expected \cite{PP}.

The order parameter $\rho$ is the number density of nearest neighbor pairs of vacant sites,
while we define the $A$ particle density as the non-order parameter $\rho_A$.
First, we perform static Monte Carlo simulations to measure the natural
density of the non-order parameter at both criticalities.
We measure the $A$ particle density $\rho_A (t,L)$, averaged over
$2\times 10^3 \sim 5\times 10^4$ survived samples for system size $L = 2^6 \sim 2^{10}$.
$\rho_A (t,L)$ relaxes to the natural density $\rho_A^{nat} \equiv \rho_A(\infty,\infty)$
in the thermodynamic limit.
At $p=p_{c1}$ and $p=p_{c2}$, we estimate $\rho_A^{nat}=0.101(1)$ and
$0.458(2)$, respectively. The distribution function for the $A$ particle density
appears to be Gaussian, so the mean density is identical to the most
probable density. We measure the most probable density in the long time limit,
which turns out to be consistent with the above mean natural density within
statistical errors.

Dynamic properties for the non-order parameter $\rho_A(t,L)$ can be extracted by studying
its temporal deviation from its steady-state value (natural density) as
\begin{equation}
\Delta\rho_A(t,L) \equiv |{\rho_A(t,L)-\rho_A^{nat}}| .
\end{equation}
As in the DP case, we assume that $\Delta\rho_A(t,L)$ follows the
same scaling behavior as the order parameter.
Using the finite-scaling theory \cite{Auk} in the steady state, $\Delta\rho_A(t,L)$
scales at criticality as
\begin{equation}
\Delta\rho_A(\infty,L) \sim L^{-\beta/{\nu_\bot}} .
\end{equation}
One can also expect the critical short time behavior as
\begin{equation}
\Delta\rho_A(t,\infty) \sim t^{-\beta/{\nu_\parallel}} ,
\end{equation}
and the characteristic time $\tau_A(L)$ scales as
\begin{equation}
\tau_A(L) \sim L^{\nu_\parallel/\nu_\bot} .
\end{equation}

From static simulations, we estimate the scaling exponents at both criticalities.
At $p=p_{c1}, p_{c2}$, we estimate $\beta/\nu_{\bot} = 0.46(2), 0.46(3)$
$\beta/\nu_{\parallel} = 0.26(2), 0.25(2)$,
and $\nu_\parallel/\nu_\bot = 1.75(5), 1.90(10)$ (see Fig.~1).
As expected, these estimations involve rather large statistical and systematic
errors, especially due to inaccuracy of the natural density values.
However, these values agree reasonably well with the DI values \cite{Jensen2},
which confirm our assumption that the non-order parameter exhibits the same type
of critical fluctuations as the order parameter.

Dynamic exponents for the order parameter in
DP systems with IMA states are known to depend on
initial conditions characterized by the non-order parameter.
To investigate this nonuniversal dynamic properties
in DI systems, we perform dynamic Monte Carlo simulations
with various initial conditions. We start with a pair of nearest
neighbor vacant sites in the absorbing background which is
controlled by the $A$ particle density $\rho_A^0$.

We measure the survival probability $P(t)$(the probability that
the system is still active at time $t$) and the
mean number of pair of vacant sites (order parameter) $N(t)$ averaged over all
samples. At criticality, these quantities scale algebraically in long time
limit as \cite{Grass2}
\begin{equation}
P(t) \sim t^{-\delta} , \ \ \ \ \ \ \ \ \ \
N(t) \sim t^{\eta} .
\end{equation}
The dynamic exponents are in general functions of initial non-order parameter density;
$\delta=\delta(\rho_A^0)$ and $\eta=\eta(\rho_A^0)$.
With initial configurations of the natural non-order parameter density,
the exponents take the ordinary DI values;
$\delta(\rho_A^{nat})=\delta_{DI}\simeq 0.285$ and $\eta(\rho_A^{nat})=\eta_{DI}\simeq 0.00$.

At $p=p_{c1}$ with $r_b=0.9$, we estimate
the values of $\delta$ and $\eta$ as $\rho_A^0$ varies from 0 to 0.35
(see Table I).
In Fig.~2, we plot the exponent shifts from the DI values ($\delta -\delta_{DI}$
and $\eta-\eta_{DI}$) versus $\rho_A^0 - \rho_A^{nat}$.
It shows a linear dependence of the exponent shifts on the deviation
of the non-order parameter from the natural density in initial configurations.
This linear dependence has been also seen in DP systems.

In summary, we investigated the IMA-IMM model in one dimension,
which show the DI-type continuous phase transition from an active phase into
an absorbing phase consisting of infinitely many absorbing states.
We show that the non-order parameter exhibits critical fluctuations
identical to the order parameter. Dynamic exponents $\delta$ and $\eta$
depends linearly on the initial non-order parameter density and
coincide with the ordinary DI values only at the natural density.


This work was supported by an Inha University research grant.

\begin{figure}
\centerline{\epsfxsize=8cm \epsfbox{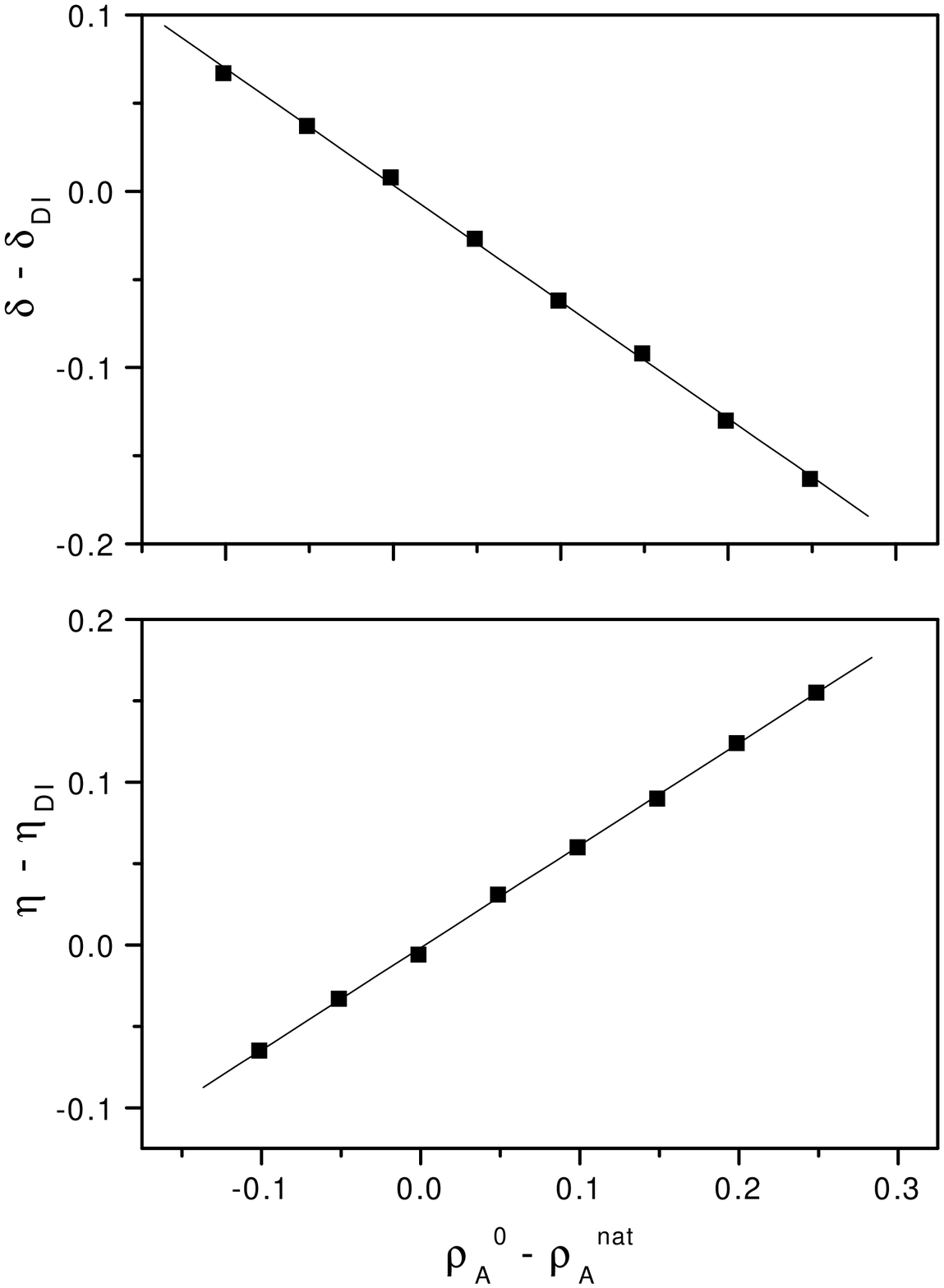}} \caption{ Temporal
dependence of $\Delta\rho_A$ and size dependence of $\Delta\rho_A$
and $\tau_A$ at $p=p_{c1}$.
The straight lines are of slope 0.26(=$\beta/\nu_{\parallel}$),
0.46(=$\beta/\nu_{\bot}$), and 1.80(=$\nu_\parallel/\nu_\bot$).}
\end{figure}

\begin{figure}
\centerline{\epsfxsize=8cm \epsfbox{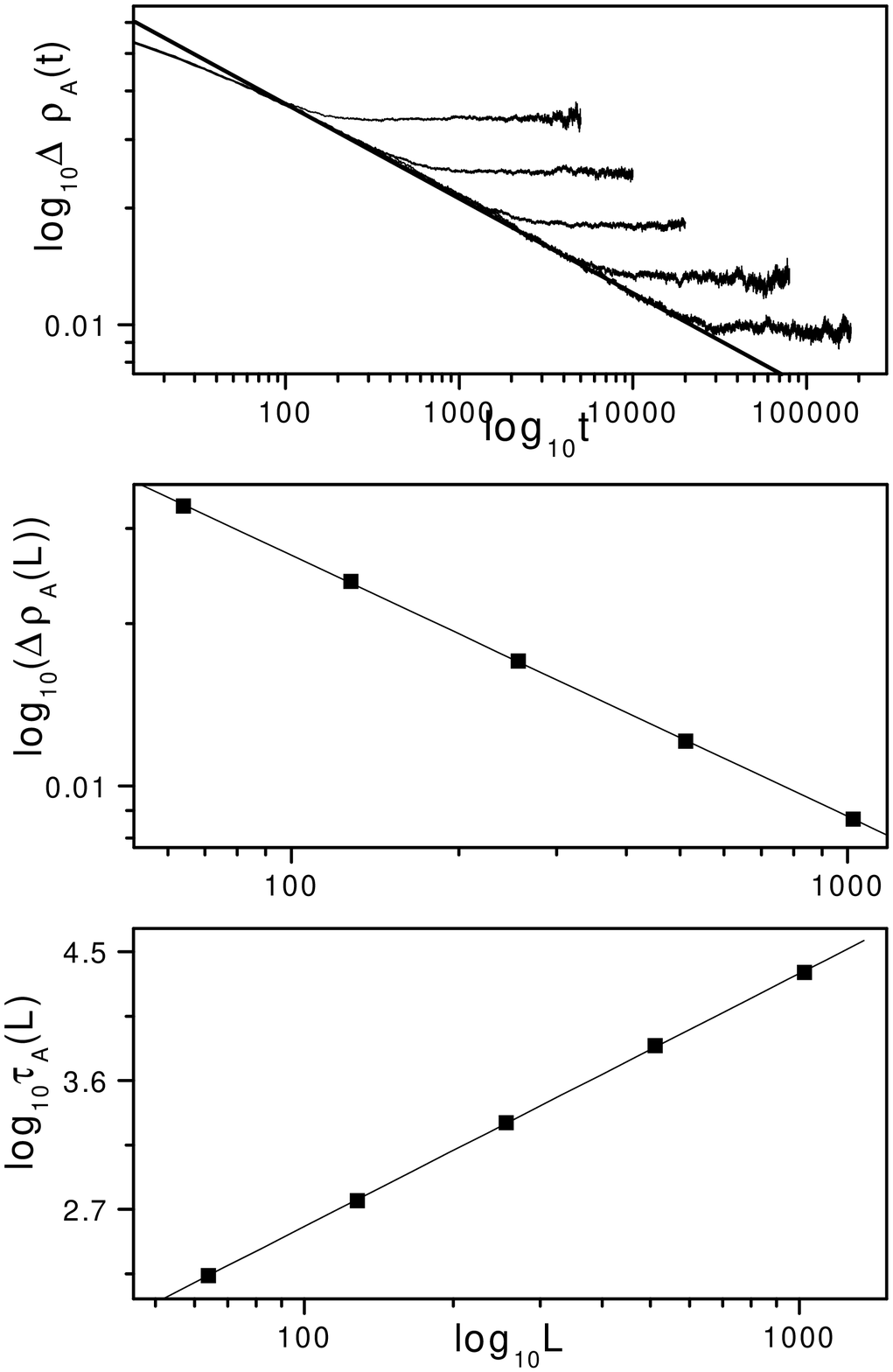}} \caption{Initial
configuration dependence of the exponents $\delta$, $\eta$ at
$p=p_{c1}$. It shows a linear dependence on the initial non-order
parameter density.}
\end{figure}

\begin{table}
\caption{Initial configuration dependence of
the dynamic exponents at $p_c = 0.1$.}

\begin{tabular}{ccc}
$\rho_{A}^0$ &  $\delta$ & $\eta$   \\ \hline $0.00$ &
$0.352(5)$ & $-0.065(10)$  \\ $0.05$ &
 $0.322(5)$ & $-0.033(7)$  \\
$0.10$ & $0.293(4)$ & $-0.006(8)$  \\ $0.15$ &
$0.258(8)$ & $0.031(9)$  \\ $0.20$ & $0.223(4)$ &
$0.060(7)$  \\ $0.25$ & $0.193(5)$ & $0.090(7)$
\\ $0.30$ & $0.155(3)$ & $0.124(6)$  \\ $0.35$ & $0.122(4)$ &
$0.155(8)$ \end{tabular}
\end{table}
\end{multicols}
\end{document}